\documentclass[aps, reprint, amsmath, amssymb, superscriptaddress]{revtex4-1}

\usepackage{amsmath}
\usepackage{amsfonts}
\usepackage{amssymb}
\usepackage{amsthm}
\usepackage{fancyhdr}
\usepackage{enumerate}
\usepackage[top=1in, bottom=1in, left=1in, right=1in]{geometry}
\usepackage{graphicx}
\usepackage[labelfont=bf]{caption}

\usepackage{color}

\usepackage{caption}
\usepackage{subcaption}

\usepackage{verbatim}

\usepackage{wasysym}
\usepackage{hyperref}

\usepackage{wrapfig}

\newcommand{\Exp}[1]{\langle #1 \rangle}

\newcommand{\yb}{ \bar y}

\begin{document}

\title{A Seascape Origin of Richards Growth}
\author{Daniel Swartz, Bertrand Ottino-L\"{o}ffler and Mehran Kardar}
\affiliation{Department of Physics, Massachusetts Institute of Technology, Cambridge,
Massachusetts 02139, USA}
\date{\today}

\begin{abstract}
First proposed as an empirical rule over half a century ago, the Richards growth equation has been frequently invoked in population modeling and pandemic forecasting.  Central to this model is the advent of a fractional exponent $\gamma$, typically fitted to the data.  While various motivations for this non-analytical form have been proposed, it is still considered foremost an empirical fitting procedure.  Here, we find that Richards-like growth laws emerge naturally from generic analytical growth rules in a distributed population, upon inclusion of {\bf (i)} migration (spatial diffusion) amongst different locales, and {\bf (ii)} stochasticity in the growth rate, also known as ``seascape noise.''  The latter leads to a wide (power-law) distribution in local population number that, while smoothened through the former, can still result in a fractional growth law for the overall population. This justification of the Richards growth law thus provides a testable connection to the distribution of constituents of the population.
\end{abstract}

\maketitle

\section{Introduction}

The mathematical description of growing populations has been an enduring topic of interest.
There is presently a great diversity of population growth models~\cite{turner1976theory, gerlee2013model, birch1999new}, which have been applied in a wide variety of contexts, 
including epidemiology~\cite{lee2020estimation, claydon2021spreading}, forestry~\cite{fekedulegn1999parameter}, developmental biology~\cite{lifeng1998richards}, cancer~\cite{gerlee2013model},  immunology~\cite{desponds2017population, gaimann2020early, amitai2017population, burnet1957modification}, and many others. 

One such model of interest is the \emph{Richards Equation}. Originally developed as an empirical model of plant development~\cite{richards1959flexible, werker1997modelling, fekedulegn1999parameter, gregorczyk1998richards}, and later used as a population growth model~\cite{pella1969generalized}, it has over the last year exploded in popularity, becoming a consistent presence in the modeling of infectious diseases, especially in COVID-19 forecasting~\cite{roosa2020short, wu2020generalized, macedo2020comparative, lee2020estimation, materassi2019some, kumar2020forecasting, wu2020generalized, roosa2020real, vasconcelos2020modelling, saif2020covid}. 

\begin{figure}[t]
\centering
\includegraphics[width = 0.5\textwidth]{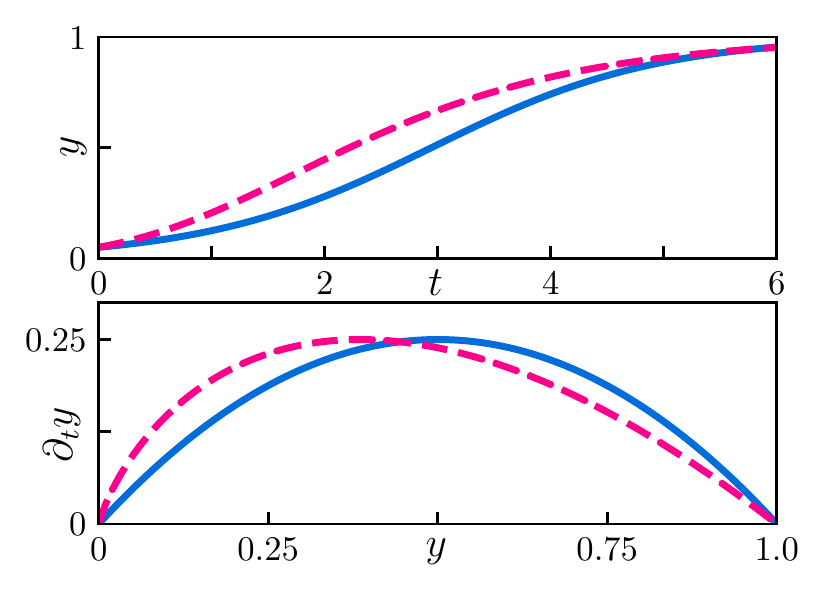}
\caption{Schematic illustration of a Richards versus logistic growth laws. The solid blue lines demonstrate a typical logistic growth curve, here with growth rate $\mu = 1$.  The dashed red lines show a Richards growth curve, here with an exponent $\gamma = 1.1$, and $a$ selected so that the saturation values are matched for illustrative purposes. (a) shows the time courses for the two equations, and (b) shows how their growth rates change with population size.  Notice how the inflection point in the Richards equation comes much earlier than that of the logistic equation. }
\label{fig:schem}
\end{figure}

The Richards model is described by 
\begin{equation} \label{eqn:richards}
\frac{d y}{d t} = \mu y - a y^{\gamma},
\end{equation}
where $y$ is a measure of the population size, $\mu$ is the initial growth rate (sometimes referred to as the population fitness), and $a$ sets the final (saturation) population to  $y_f = (\mu/a)^{1\over \gamma-1}$. {The feature that distinguishes} this from the celebrated and more natural Verhulst logistic equation~\cite{verhulst1838notice} is the non-analytic (shape) parameter $\gamma > 1$. When $\gamma = 2$, a logistic equation is retrieved, but for any other value, the inflection point of the growth curve becomes off-center, leading to asymmetric growth curves, as shown in Fig.~\ref{fig:schem}. 

This shape parameter $\gamma$ indeed gives the Richards equation character, but its meaning beyond a fitting device is mysterious. Because $\gamma$  takes on factional values, this makes the dynamics of Richards growth nonanalytic, making its origins theoretically nontrivial. Although there have been proposed derivations of Richards-like growth, they have either leaned on an underlying fractal topology~\cite{mombach2002mean, martinez2008continuous, materassi2019some}, or relied on a detailed  manipulation of an SIR model, which may not be robust under model perturbation~\cite{wang2012richards, macedo2020comparative}.  In practice, Richards growth is still  considered an empirical law and lacks a fully universal origin, or a physically intuitive interpretation of the shape parameter $\gamma$. 

In this paper, we propose a plausible origin for emergence of Richards-like growth in  distributed populations from generic local analytical forms. Specifically, we shall use the \emph{Fisher Equation} as  starting point. The deterministic Fisher equation~\cite{fisher1937wave, aronson1978multidimensional, gourley2000travelling, kwapisz2000uniqueness, hallatschek2009fisher} is one of the most basic models of spatial population growth, written as 
\begin{equation}\label{eqn:fisher}
\frac{dy(x,t)}{dt} = \mu y - a y^2 + D \nabla^2 y~.
\end{equation}
This is distinguished from a logistic equation by the presence of a spatial coordinate $x$ and a diffusion term $D$ which sets the rate of local migration. 

The deterministic Eq.~\eqref{eqn:fisher} should in principle include reproductive stochasticity. 
While there are many kinds of randomness that may influence population growth (e.g., demographic noise, with amplitude proportional to $\sqrt{y}$~\cite{durrett1994importance,butler2009robust, ovaskainen2010stochastic,traulsen2012stochastic,constable2016demographic,weissmann2018simulation}), we are interested in what is called \emph{Seascape Noise}~\cite{merrell1994adaptive, mustonen2009fitness, mustonen2010fitness, agarwala2019adaptive, iram2019controlling, trubenova2019surfing, vincenzi2014extinction, cvijovic2015fate, peischl2012establishment, munoz2003multiplicative}. This noise arises from observing that the fitness $\mu$ of a population is not necessarily a static, uniform quantity. Population fitness can vary based on minor environmental conditions, access to resources, microscopic mutation rates, and other such factors. So we generalize from a static fitness landscape $\mu(x)$ to one that varies in time $\mu(x,t)$, much as the sea surface changes in time. 

By introducing seascape noise into the Fisher equation, we obtain the \emph{Seascape Fisher Equation}
\begin{equation} \label{eqn:seascape_fisher}
dy = \left(\mu y - ay^2 + D \nabla^2 y\right)dt + \sigma y dW,
\end{equation}
with $\sigma^2$ being the variance of the fitness noise, and $dW = dW(x,t)$ is an uncorrelated Wiener process.  While our noise will be uncorrelated, we should note certain applications may wish to consider more spatially correlated environmental variations~\cite{moran1953statistical, moran1953statistical2, ranta1997moran, liebhold2004spatial}. Moreover, we consider this stochastic differential equation under an It\^{o} interpretation, since its assumptions pair well with the fact that generations in a popuation are discrete, though interesting seascape results exist for alternate interpretations~\cite{ao2007existence, tang2014controlling, kwon2011nonequilibrium}.  

Since both the Fisher equation and seascape noise are very versatile concepts, this makes their combination in Eq.~\eqref{eqn:seascape_fisher} similarly adaptable. Variants of this much-studied model~\cite{grinstein1996phase, tu1997systems, munoz1998nonlinear, hinrichsen2000non} have connections to the Kardar-Parisi-Zhang equation in the field of surface growth problems~\cite{kardar1986dynamic, kardar1987scaling, derrida1990directed, chu2016probability, borodin2014free, medina1989burgers, quastel2015one, sasamoto2010one, halpin2013extremal}, the study of directed polymers in random media~\cite{grinstein1996phase, tu1997systems, munoz1998nonlinear, hinrichsen2000non}, symmetry breaking~\cite{van1994mean, van1994noise, van1997nonequilibrium}, theoretical ecology~\cite{hakoyama2005extinction, ovaskainen2010stochastic, melbinger2015impact, xu2014two, pearce2020stabilization, chotibut2017striated}, immunology~\cite{desponds2016fluctuating}, and economics~\cite{bouchaud2000wealth}. 

In this work, we argue that spatially averaging the seascape Fisher equation is sufficient to produce Richards-like growth curves. 
In Sec.~\ref{sec:mean_field} we introduce a discretization of the model, replacing $y(x)$ with $y_i$ for locations (nodes) $i=1,\ldots, N$. In the context of this model diffusion between neighboring sites can be replaced with migration between any pair of sites at rates $M_{ij}$. 
A mean-field limit is then obtained if migration rates are equal between any pair of sites
($M_{ij}=D/N$). This limit was considered by us \cite{ottino2020population} in the context of population
extinction, and similar models by others in different contexts~\cite{van1994mean, van1994noise, lombardo2014nonmonotonic, gomes2018mean, desponds2016fluctuating}. 
The steady steady (long-time) limit of the probability distribution $\rho(y)$ can be obtained
exactly in this mean-field limit, and characterized by a power-law distribution. 
A wide power-law distribution, characterized by non-analytic dependencies of its moments,
suggests a natural route for obtaining Richards-like growth.
Unfortunately, we could not solve the full dynamic behavior of the model, even in the mean-field
limit. 
As an alternative, in Sec.~\ref{sec:seasonal} we introduce a model which alternates the linear
and nonlinear parts of the model. This {\it seasonal} growth model retains the features of
seascape stochasticity and migration which we believe are the cause of Richards growth,
and (with some assumptions) is exactly solvable;  numerical simulations confirm these expectations.
In Sec.~\ref{sec:other} we consider a number of extensions:
providing a procedure for testing the model by examining the dependence of the second moment on the mean
(Sec.~\ref{sec:seasonalgrowth});
indicating the universality of the results for generalized growth equations (Sec.~\ref{sec:universal});
numerical studies of  the seascape Fisher equation in one and two dimensions (Sec.~\ref{sec:SFE});
and finally indicating the appearance of Gompertz growth in the strong noise limit (Sec.~\ref{sec:Gompertz}).
The concluding Sec.~\ref{sec:discussion} provides an overview, and outlook for future studies.

\section{Mean-Field Steady State} \label{sec:mean_field}
The intuition for why seascape noise is a viable mechanism for Richards growth comes from 
earlier studies of a mean-field version of the problem. 
We begin with the full seascape Fisher equation Eq.~\eqref{eqn:seascape_fisher} in $d$ spatial dimensions. 
We discretize space into $N$ lattice sites with unit spacing indexed by $i = 1,2,\ldots,N-1$ with periodic boundary conditions. 
The field $y(x,t)$ is replaced by a single value at each lattice site, $y(x = i,t) \rightarrow y_i(t)$. 
The Laplacian on this lattice takes the form of a migration rate $M_{ij}$ giving the rate of migration from site $j$ to $i$. 
For example, on a regular one-dimensional lattice, $M_{ij} = \delta_{i+1,j} + \delta_{i-1,j} - 2\delta_{ij}$.
The discretized version of the stochastic Fisher equation now takes the form 
\begin{equation}
    dy_i = (\mu y_i - a y_i^2 + \sum_j M_{ij} y_j)dt + \sigma y_i dW_i\,.
\end{equation}
One can use the above equation and generalize the matrix $M_{ij}$ to any migration connectivity network. We then proceed to take the mean field limit by using a complete graph, defining $M_{ij} = D/N$, corresponding to a population that can migrate with ``steps" of arbitrary length.

The stochastic Fisher equation in the mean-field case now takes the simple form
\begin{equation} \label{eqn:mean_field}
    dy_i = (\mu y_i - a y_i^2 + D(\bar{y} - y_i))dt + \sigma y_i dW_i\,,
\end{equation}
where $\bar{y}$ is the spatial average of $y$\footnotetext[1]{Because $\yb$ depends on the solution $\rho$, this makes the SDE~\eqref{eqn:mean_field} a McKean-Vlasov equation~\cite{mckean1966class, sznitman1991topics, graham1992mckean, mezerdi2019approximation, mishura2016existence}}~\cite{Note1}. 
The steady state (long-time) behavior of this stochastic ordinary differential equation 
is obtained as the stationary solution of a corresponding Fokker--Planck equation 
(see Ref.~\cite{ottino2020population} and Sec.~\ref{sec:seasonalrichards}), and is proportional to
\begin{equation} \label{eqn:mfsteadystate}
    \rho(y) = e^{-c_D \bar{y} / y} y^{-2-c_D + c_\mu} e^{-c_a y}\,,
\end{equation}
where $c_D = 2 D / \sigma^2, c_\mu = 2 \mu /\sigma^2, c_a = 2 a / \sigma^2$. In the limit $N \rightarrow \infty$ we can identify $\bar{y} = \Exp{y}$ as a parameter to be found self-consistently~\cite{ottino2020population, van1994mean, van1994noise, lombardo2014nonmonotonic, gomes2018mean,munoz2005mean}. 
The steady state distribution thus has a power-law with upper and lower cutoffs of $c_a^{-1}$ and $c_D \bar{y}$ respectively. Distributions of this form give rise to \emph{anomalous scaling} in the moments, which we conjecture as the mechanism for  Richards growth.   

Demanding that $\bar{y}$ be the mean of this distribution implies                                     the self consistency condition
\begin{equation} \label{eqn:self_consistent}
    \bar{y} = \frac{\int y \rho(y)\, dy}{\int \rho(y)\, dy} \, .
\end{equation}
By manipulating this condition~\cite{ottino2020population}, we arrive at a moment scaling relationship of 
\begin{equation}
    \Exp{y^2} \propto \Exp{y}^{1+\min(c_D-c_\mu, 1)}.
\end{equation}
Thus, for $c_D-c_\mu < 1$ we have that the second moment scales with the first moment as $\Exp{y^2} \sim \bar{y}^{1+c_D-c_\mu}$, while for $c_D-c_\mu > 1$, we have a more typical scaling $\Exp{y^2} \sim \bar{y}^2$. This anomalous scaling \textit{in steady state} is what motivated our hypothesis for Richards growth.
However, absent a solution of the time-dependent Fokker-Planck equation, we cannot identify parameter regimes where the previous scaling holds away from steady state. 
As such, we introduce a seasonal model  that allows us to take advantage of the results of this section. 

\begin{figure}[t]
\centering
\includegraphics[width = 0.5\textwidth]{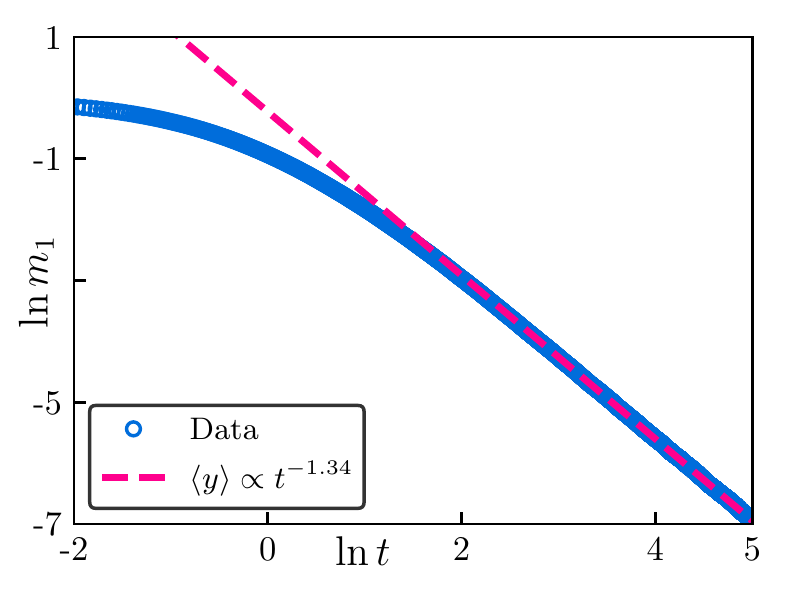}
\caption{Decay of mean field model Eq.~\eqref{eqn:mean_field} in the case of $\mu = 0$.   The blue circles show the numerically simulated decay of the mean, here with $a = 1$, $D = 0.8$, $\sigma^2 = 2$, $ N = 2^{20}$, and $dt = 0.01$.  The dashed red line shows a fitted power law to the tail of the decay. Notice that the fitted power law is around -1.34, and is slightly different than the predicted value of $-1/c_D = -1.25$.}
\label{fig:decay}
\end{figure}

A justification for employing the steady-state solution in a dynamic setting comes from the case of population decay for $\mu=0$:  
If we take the scaling, $\Exp{y^2} \propto \Exp{y}^{1+\min(c_D,1)}$, 
from the stationary-state,
and  assume that it is maintained quasi-statically during the decay process, we can make a prediction for the decay of the mean.
In particular, $\partial_t \Exp{y} = -a\Exp{y^2} \propto -\Exp{y}^{1+c_D}$,
indicates that the mean should scale as $t^{-1/c_D}$.  
The numerical results plotted in Fig.~\ref{fig:decay},
indicate a power-law decay roughly consistent with the this quasi-static approximation.  
A similar argument is made in Ref.~\cite{dornic2005integration}
where power-law decays as in Fig.~\ref{fig:decay} are demonstrated.
To avoid problems with the quasi-static assumption during growth, we next introduce a two-state model next that allows to more controllably take advantage of this solution.

\section{Seasonal growth model}\label{sec:seasonal}
The combination of nonlinearity and stochasticity complicates the study of Eqs.~\eqref{eqn:seascape_fisher} and \eqref{eqn:mean_field}.
To simplify analysis while maintaining the qualitative features we believe to be responsible for the Richards form, we separate non-linear growth and stochastic migration. 
 To do so, we introduce a seasonal growth model composed of two distinct stages:
\begin{enumerate}
    \item \emph{Exploration}: The exploration phase (season) has the population diffusing in the
    presence of seascape noise, as described by 
     \begin{equation*}
        dy = D\nabla^2 y dt + \sigma y dW\,.
    \end{equation*}
In the mean-field limit, $D\nabla^2 y$ is replaced by $D(\bar{y} - y)$, where $\bar{y} = \Exp{y}$ at the start of the exploration phase. 
The exploration phase is run for a time $T_e$ with $DT_e \gg 1$.

As these dynamics conserve the mean,  $\bar{y}$ can be treated as a constant. 
In the absence of reproduction in the exploratory phase, a new interpretation for
seascape noise is necessary. A potential cause is random extinctions of the populations
at different locales and times. In this interpretation, the noise must have a negative mean,
and $\Exp{y}$ will decrease through the process. 
For a large enough population (to avoid the possibility of extinction) this overall loss can be 
restored in the subsequent growth stage, without qualitatively changing the results.

    \item \emph{Growth}: 
 Even when starting with a uniform distribution of numbers at each node, the stochasticity in the
 exploratory phase leads to a broad distribution of $y_i$ at the end of this interval.
 In the subsequent growth phase, at each node we implement reproductive growth following the logistic equation,
        \begin{equation*}\label{eqn:logistic}
        \partial_t {y_i} = \mu y_i - a y_i^2\,,
    \end{equation*}
    for a short time $T_g$ where $\mu T_g \ll 1$.
    
   Equation~\eqref{eqn:logistic} is deterministic, and can easily be solved to give the final population at each node in terms of the initial value. 
    The exact form of the equation governing growth and saturation is not important, and as discussed in the following, any analytical form leads to similar results.
\end{enumerate}

We alternate between exploration and growth for a large number of times (seasons) to generate trajectories for the moments of the population $m_n(t) = \Exp{y(t)^n}$.  

\subsection{Stochastic exploration} \label{sec:seasonalrichards}
In the exploration phase of the mean-field model, the distribution of $y$ evolves according to the Fokker-Planck equation
\begin{equation}
    \partial_t \rho = -\partial_y \left[D(\bar{y} - y) - \frac{\sigma^2}{2}\partial_y(y^2 \rho) \right].
\end{equation}
The stationary steady state solution is found by demanding a vanishing probability flux, and is proportional to 
\begin{equation}
    \rho(y) = e^{-c_D \bar{y} / y} y^{-2-c_D}\,,
\end{equation}
which has the advertised structure of a power-law with $c_D=2D/\sigma^2$, and
a lower cutoff set by $c_D \bar{y}$. 
Power laws arising from models with seascape noise are quite common, so this is well-precedented~\cite{leigh1981average, lande1993risks, ovaskainen2010stochastic, desponds2016fluctuating, desponds2017population, barghathi2017extinction, foley1994predicting}. {However, the continuous variation of the exponent
with the ratio $c_D$ is unusual and likely a feature of the mean-field limit;
 though some similarly continuous non-universal exponents have also appeared in directed percolation under temporal disorder~\cite{jensen1996temporally, jensen2005low}.} 
 
 However, the steady state solution is only valid in the limit $T_e \rightarrow \infty$; for any finite $T_e$, the power-law tail  won't extend to infinity, and higher moments will not diverge (see Appendix~\ref{appendix:seascapediff_moment}).  In addition, expansion and growth occur simultaneously in the original seascape Fisher equation, making the establishment of the full power-law tail implausible.  Because of this, we also impose an upper cutoff $\Lambda$ to this distribution, simplifying it to
 \begin{equation}
    \rho(y) \propto y^{-2-c_D}\quad{\rm for}\quad \Upsilon< y < \Lambda\,,
    \label{eqn:pareto}
\end{equation}
with $\rho(y) = 0$ otherwise.  
We have additionally imposed a lower cutoff $\Upsilon$ to ensure that the mean is equal to $\bar{y}$.
For $0<c_D < 1$, and as long as $\Lambda\gg \Upsilon$, the first  moment of the distribution is given by
\begin{equation}\label{eqn:meanPareto}
    \Exp{y} \approx \frac{1+c_D}{c_D}\Upsilon\,,
\end{equation}
setting the lower cutoff to
\begin{equation}
\Upsilon=\frac{c_D}{1+c_D}\bar{y}\,.
\end{equation}
The second moment is now given by
\begin{equation}\label{eqn:mom2Pareto}
    \Exp{y^2} \approx  \frac{1+c_D}{1-c_D}\Lambda^{1-c_D}\Upsilon^{1+c_D} \propto \Exp{y}^{1+c_D}.  
\end{equation}

\subsection{Deterministic growth} \label{sec:seasonalgrowth}

Once the growth phase begins, the variable $y_i$ of nodes, distributed according to Eq.~\eqref{eqn:pareto}, serve as initial conditions  for the logistic equation. 
For each node, the solution of the logistic equation gives
\begin{equation}
   {y_i(t)} = \frac{e^{\mu t} \mu y_i}{ay_i(e^{\mu t}-1) + \mu} \,. 
\end{equation}
Averaging over the initial conditions now leads to
\begin{equation}
    \Exp{y(t)} = \frac{\int^\Lambda_{\Upsilon } \frac{e^{\mu t} \mu y}{ay(e^{\mu t}-1) + \mu} \rho(y) dy}{\int_{\Upsilon }^\Lambda \rho(y) dy}\,. 
\end{equation}
While this integral can be approximated in detail (see Appendix \ref{appendix:twophase_moment}), as long as the growth phase only occurs for a small time ($\mu T_g \ll 1$) we can expand our result for the mean at the end of the growth phase to $O(T_g)$ and find 
\begin{alignat}{1}
    \Exp{y(t+\Delta t)}  =~ & (1+  \mu\Delta t) \Exp{y(t)} - a \Exp{y^2(t)}\Delta t\,. 
\end{alignat}
If we take advantage of the fact that $\Lambda \gg \Upsilon $, then we can take advantage of the
previously computed moments to obtain
\begin{equation} \label{eqn:DeterministicRichards}
    \frac{\Delta \Exp{y(t)}}{\Delta t} = \mu \Exp{y}\left[1-  \frac{c_D}{1-c_D}\frac{ a\Lambda}{\mu}\left(c_D\Exp{y}\over (1+c_D)\Lambda\right)^{c_D}\right]\,.
\end{equation}
Therefore, when we write the dynamics of the mean in the growth phase, we have one term that scales as $\Exp{y}$ and one term that scales as $\Exp{y}^{1+c_D}$, leading to non-analytic, Richards-like growth.  

\begin{figure}
    \centering
    \includegraphics[width = 0.5\textwidth]{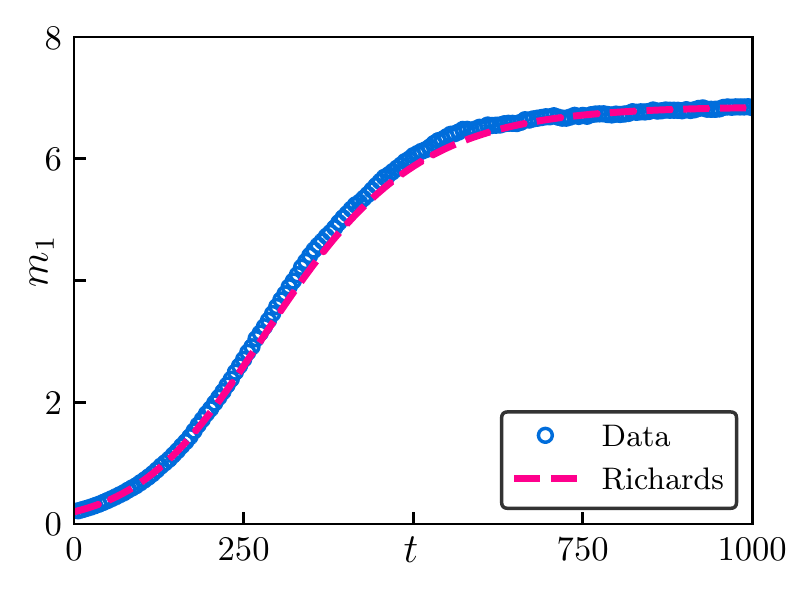}
    \caption{Simulation of the seasonal growth model.  Blue circles show the results of numerically solving the model over 175 cycles of exploration and growth for $N = 2^{20}$ nodes. The dashed line is a solution to the Richards equation with growth rate $\tilde{\mu}$ and exponent $c_D$ as calculated in the main text, and with a numerically found carrying capacity. The simulation used $\mu =  1$, $a = 10^{-3}$, $\sigma = \sqrt{2}$, $D = 0.5 $, $\Lambda = 500$, $T_g = 0.1$, and $T_e = 6$.}
    \label{fig:stairstep}. 
\end{figure}

Upon iterating these dynamics at times $t_j = j(T_e + T_g)$, we find a population trajectory $m_1(t_j) = \Exp{y(t_j)}$ governed by approximate Richards law to leading order in time and $\bar{y}/\Lambda$, of
\begin{equation} \label{eqn:richardsodetwophase}
    \frac{dm_1(t)}{dt} = \tilde{\mu} m_1(t) - \tilde{a}(\Lambda) m_1(t)^{1+c_D} + O({m_1}^2),
\end{equation}
where $\tilde{\mu} = \frac{1}{1+T_e/T_g} \mu$ is the effective growth rate and is cutoff-independent. The carrying capacity is set by $\tilde{a}(\Lambda)$ which is some function of the cutoff and can be found numerically. We confirm these results in Fig.~\ref{fig:stairstep} where we see excellent agreement between the simulated mean population and the analytical prediction. We also note that the numerical results can also be reasonably fitted to a logistic sigmoid. However, what is significant is that models such as Eq.~\eqref{eqn:seascape_fisher} contain the essential ingredients to analytically justify a Richards growth law.

\section{General considerations} \label{sec:other}
In this section we test the generality and applicability of the results from the seasonal model
in a number of other contexts.

\subsection{Dynamics of Seascape Mean-field} \label{sec:seasonalgrowth}
In the seasonal model the non-analytic form of Richards growth emerged from the power-law character
of the distribution of local numbers, established from the combination of migration and seascape noise
in the exploration phase. 
In the original model, and many likely realizations, growth and migration occur simultaneously,
preventing formation of a well defined quasi-stationary power-law distribution.
This complicated testing the foundation of our explanation of Richard's growth in terms of 
a relation between the growth exponent of the distributed population and its distribution amongst
different locations.

\begin{figure}
    \centering
    \includegraphics[width = 0.5\textwidth]{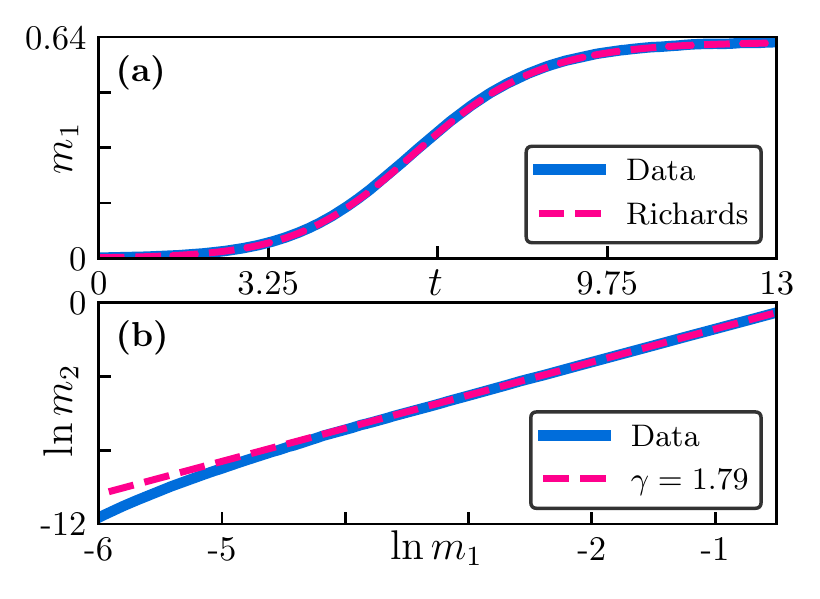}
    \caption{Numerical simulation of the mean-field dynamics of Eq.~\eqref{eqn:mean_field},
    for $N = 2^{20}$ nodes with $\mu = a = 1$, $\sigma = \sqrt{2}$, $D = 1.2$, and $dt = 1/120$.
    The sigmoid curve in (a) is well fitted to a Richards equation with shape parameter $\gamma=1.79$.
    (Only times after $4/\mu$ were used to fit the Richards exponent.)
    Figure (b)  shows the relation between the first and second moments of the population on a log scale. 
   The dashed line shows the best fit power-law $m_2 \propto{m_1}^\gamma$.
   }
    \label{fig:meanfield}
\end{figure}

To address these questions we simulated the full dynamics of Eq.~\eqref{eqn:mean_field} for $N=2^{20}$
nodes. 
The mean, $\Exp{y} = m_1$, of this distributed population grows in sigmoid fashion as depicted in Fig.~\ref{fig:meanfield}a.
In the absence of noise, with $\mu=a=1$, the population would have saturated to $\mu/a=1$;
seascape noise reduces the saturation value to roughly 0.64. 
Allowing for a variable Richards growth exponent, the sigmoid curve is {well-}fitted with a value of $\gamma=1.79$.
(Note that with the simultaneous action of growth and diffusion, we do not expect $\gamma=1+c_D$,
with $c_D=1.2$ for the chosen parameters of $\sigma^2=2$ and $D=1.2$.)  

Nonetheless, we can test whether something like our proposed mechanism is at work by plotting
the time evolution of the second second moment $\Exp{y^2} = m_2$ as a function of the first $\Exp{y} = m_1$.
As depicted in Fig.~\ref{fig:meanfield}b, at intermediate and late times, we find  $m_2 \propto{m_2}^\gamma$--
a straight line in the log-log plot with a slope of $\gamma=1.79$.   {Such log-log fits are a convenient way to retrieve the growth exponent $\gamma$ in models such as these, and will be used throughout this paper.} (The intercept for the log-log line 
is also useful, and related to the carrying capacity.)

Finding the first and second moments of a distributed population in a natural setting should be easier than  
characterizing the whole distribution. A relation such as  $m_2 \propto{m_2}^\gamma$, coupled with a Richards
fit to the growth curve with the same exponent would provide a good test of the proposed hypothesis.

\subsection{Universality}
\label{sec:universal}
The formulation of the model in terms of the logistic equation may give the impression that the obtained
results are a consequence of its quadratic form.
There are in fact many other analytic growth functions with saturation that generate sigmoid curves.
The anomalous scaling, induced by the heavy-tailed distributions attributed to spatial diffusion and seascape noise, actually confers a universality to these results, independent of the assumed growth equation.
Let us consider a generic {\it analytical} growth curve $g(y)$ with an initial growth rate $\mu$ that saturates
to a  carrying capacity $y=K$, which can be written as
\begin{equation}
    g(y) = \mu y \left(1-\frac{y}{K}\right)f(y)\,.
    \label{eqn:localgrowth}
\end{equation}
The analytic function $f(y)$, with $f(0) = 1$, alters the growth away from the purely logistic case, 
 and we will demand that $f(y) > 0$ for all $y$ to ensure the only fixed points of the local growth are 
 at $y = 0$ and $y=K$. This also guarantees that our local dynamics given $g(y)$ only have simple (non-repeated) roots. 
 
We will again demonstrate how such a generalized model behaves in the seasonal dynamics introduced before, where
the population alternates {dispersal with} seascape noise, {and local growth according to} Eq.~\eqref{eqn:localgrowth}. As before, we restrict analysis to mean-field dispersion during the exploration phase,
leading to 
\begin{equation}
    \begin{cases}
    dy_i = D(\bar{y}-y_i)dt + \sigma y dW_i(t) & \text{Exploration}\\ 
    dy_i = \mu y_i (1-y_i/K)f(y_i)dt & \text{Growth}
    \end{cases}\,.
\end{equation}
The exploration phase is again assumed to be long enough to establish a power-law distribution for local numbers, as in Eq.~\eqref{eqn:pareto}.
With the mean set by $\Exp{y}=\bar{y}\propto\Upsilon$,  (for $\Lambda\gg\Upsilon$) the higher moments behave as 
\begin{equation}
\Exp{y^n} \simeq \frac{ \Lambda^{n-1-c_D}}{n-1-c_D}c_D\Upsilon^{1+c_D} \propto \Lambda^{n-1-c_D}\Exp{y}^{1+c_D}\,.
\end{equation}
Notably, each higher moment has the same scaling with $\bar{y}$, although the prefactors vary by powers 
of the upper cutoff $\Lambda$.

\begin{figure}
    \centering
    \includegraphics[width = 0.5\textwidth]{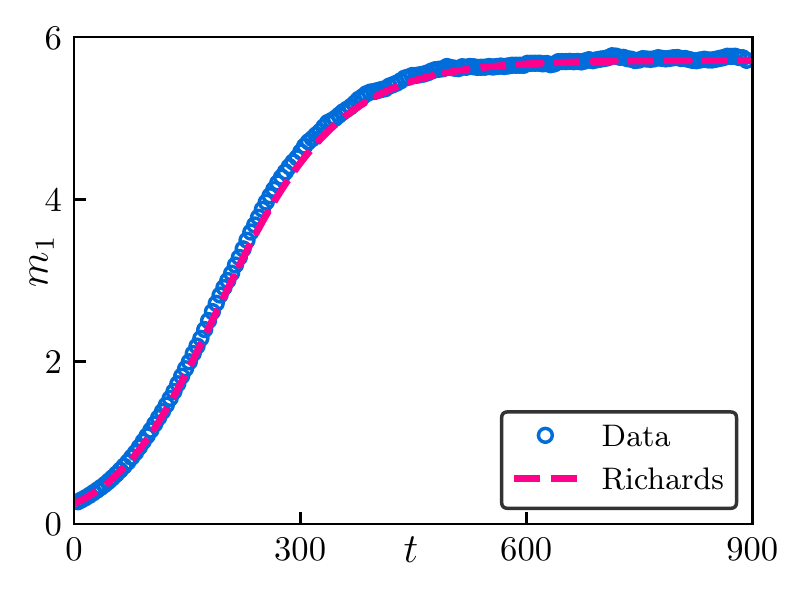}
    \caption{Numerical simulation of seasonal growth dynamics with a deterministic phase of Eq.~\eqref{eqn:localgrowth},
    with $f(y)=1+ 0.2 y/K + 0.3 (y/K)^2$.
    The blue circles show the value of the average population at the end of each cycle, and the red dashed line shows the solution to Eq.~\eqref{eqn:richardsodetwophase} with $\tilde{\mu} = \frac{\mu}{1+T_e/T_g}$. We take the carrying capacity from the numerics in order to find $\tilde{a}(\Lambda)$. The simulation used $N = 2^{20}$, $\mu =  1$, $a = 10^{-3}$, $\sigma = \sqrt{2}$, $D = 0.6 $, $\Lambda = 200$, $T_g = 0.1$,  and $T_e = 5$.}
    \label{fig:universal}
\end{figure}

Following a localized growth step of duration $\Delta t\ll \mu^{-1}$, the mean population is increased
by $\Delta t\Exp{g(y)}$. Since $g(y)$ can be expanded as a power series $ g(y) = \mu y +g_2 y^2 + g_3 y^3+\cdots$, we obtain
\begin{equation}
 \Exp{g(y)}=\mu \Exp{y}-A(\Lambda)\Exp{y}^{1+c_D}+{\cal O}(\Exp{y}^{2},\Exp{y}^{2+c_D})\,.
    \label{eqn:analyticexpansion}
\end{equation}
Once again,  averaging over the exploration distribution  establishes a Richards growth law;
the difference with the purely logistic case is absorbed into the (cut-off dependent) coefficient of
the non-analytic saturation term.
We test this universality by simulating seasonal growth with a quadratic $f(y)$ curve in Fig.~\ref{fig:universal}.

\subsection{Seascape Fisher Equation in 1 and 2D}
\label{sec:SFE}

\begin{figure}
    \centering
    \includegraphics[width = 0.5\textwidth]{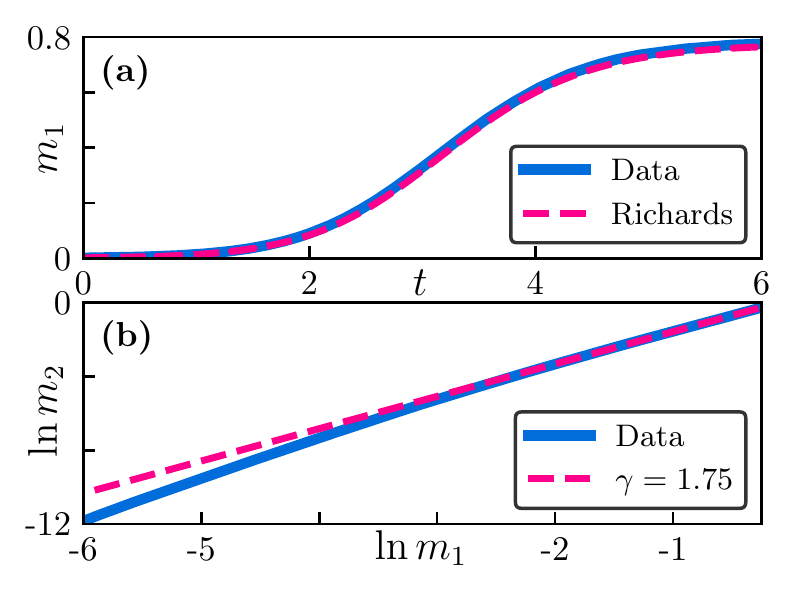}
    \caption{Emergence of Richards growth in simulations of the seascape Fisher  Eq.~\eqref{eqn:seascape_fisher}
    in one dimension.  Figure (a) compares the simulated first moment to the solution of a Richards equation, with a numerically fitted shape parameter $\gamma$. Figure (b) shows the relation between the first and second moments on a log scale. The dashed line show the best fit power-law $m_2 \propto{m_2}^\gamma$. We simulate $N = 2^{20}$ nodes with $\mu = a = D = 2$, $\sigma = \sqrt{2}$,and $dt = 1/200$. Only times after $4/\mu$ were used to fit the Richards exponent.}
    \label{fig:onedim}
\end{figure}

\begin{figure}
    \centering
    \includegraphics[width = 0.5\textwidth]{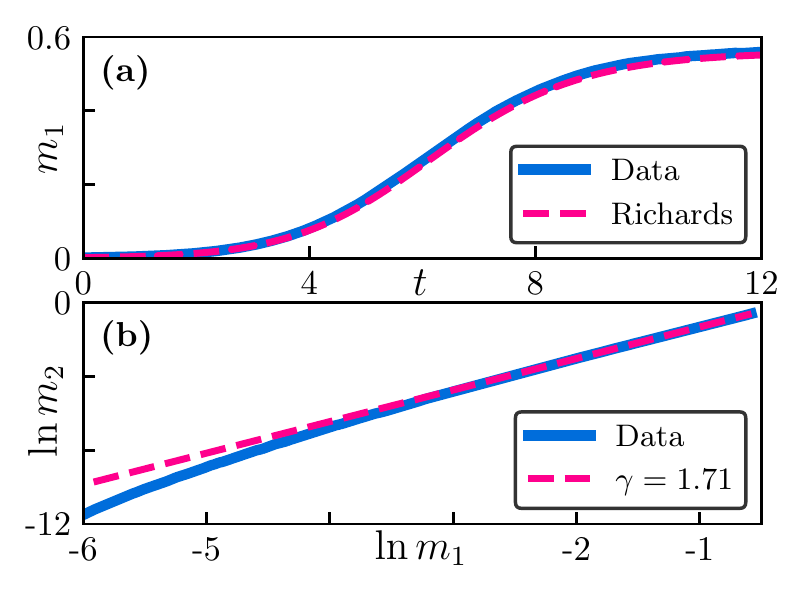}
    \caption{Emergence of Richards growth in simulations of the seascape Fisher  Eq.~\eqref{eqn:seascape_fisher}
    in two dimensions.   Figure (a) compares the simulated first moment to the solution to the Richards equation, with the shape parameter $\gamma$ determined numerically. Figure (b) shows the relation between the first and second moments on a log scale. The dashed line show the best fit power-law $m_2 \propto{m_2}^\gamma$. We simulate $N = 2^{10}$ nodes with $\mu = a = D = 1$, $\sigma = 2$,and $dt = 1/100$. Only times after $4/\mu$ were used to fit the Richards exponent.}
    \label{fig:twodim}
\end{figure}

Armed with insights from the mean-field model, we now return to the seascape Fisher Eq.~\eqref{eqn:seascape_fisher}.
To do so,  we numerically simulate Eq.~\eqref{eqn:seascape_fisher} using a central-difference discretization 
of the Laplacian on a lattice. 
We observe that the mean population can be well fit by a Richards curve, with the Richards exponent and 
carrying capacity are numerically determined from the data as in the mean-field case (Sec.~\ref{sec:seasonalgrowth}).
As depicted in  Fig.~\ref{fig:onedim} this procedure leads to a characterization of the one dimensional growth
with an emergent Richards shape parameter of $\gamma\approx 1.75$.
The corresponding results in two dimensions (for a square lattice and periodic boundary conditions)
are presented in Fig.~\ref{fig:twodim} with shape parameter $\gamma\approx 1.71$.
Here, we present these shape parameters as effective exponents describing particular simulations.
Further investigation, possibly following Refs.~\cite{dornic2005integration,munoz1998nonlinear,genovese1999recent}
may shed light on the values of these exponent and their universality.

\subsection{Gompertz growth law for strong noise} \label{sec:Gompertz}
We can rewrite Eq.~\eqref{eqn:DeterministicRichards}   (see also Eq.~\eqref{eqn:analyticexpansion})
as
\begin{equation} \label{eqn:RechardsCapacity}
    \frac{\Delta \Exp{y(t)}}{\Delta t} = \Exp{g(y)}=\mu \Exp{y}\left[1- \left(\Exp{y}\over K\right)^{c_D}\right]\,,
\end{equation}
in terms of a final capacity $K$. In the limit of strong seascape noise, {we have $c_D=2D/\sigma^2\to0$, and if we let $\mu = \tilde \mu /c_D$, then}  Eq.~\eqref{eqn:RechardsCapacity} acquires the form
\begin{equation} \label{eqn:Gompertz}
    \frac{\Delta \Exp{y(t)}}{\Delta t} =-\tilde\mu \Exp{y}\ln \left(\Exp{y}\over K\right)\,.
\end{equation}
The above form, known as the {\emph Gompertz equation} is another commonly
used growth law~\cite{gerlee2013model, turner1976theory, werker1997modelling}. 
The qualitative prediction of the model is that the Gompertz law should emerge in systems subject
to strong seascape noise.

\section{Discussion} \label{sec:discussion}

Growth and saturation phenomena are prevalent in nature, leads a host of models and mathematical
forms to quantify their description. The Richards and Gompertz laws stand prominently by providing
successful empirical fits to sigmoid curves in diverse contexts~\cite{gerlee2013model, turner1976theory, werker1997modelling, pella1969generalized, martinez2008continuous, lee2020estimation, werker1997modelling, materassi2019some, gregorczyk1998richards, kumar2020forecasting, roosa2020real, lifeng1998richards, fekedulegn1999parameter, richards1959flexible}.
A typical application involves data obtained by summing (averaging) numbers from a distributed dataset.
The appearance and success of such non-analytical forms in describing sums is quite surprising:
In the same sense that the probability distribution for the sum of many variables typically assumes
the analytic Gaussian form, the most natural time evolution for the sum is an analytic growth law
(whose first two terms form the logistic equation).
A non-analytic series (much like a critical exponent in critical phenomena) requires specific justification.

In this work, we argue that there is good theoretical and numerical evidence to suggest that Richards-like growth 
emerges as a natural consequence of the combination of dispersion and (seascape) stochasticity in a
large distributed population.  In steady state, the combination of the two establishes a broad power-law
distribution at distinct locales; averaging any analytic growth rule with such a distribution leads to a 
non-analytic Richards form. This argument, which nicely connects distribution of local numbers to the time
evolution of the mean, however relies on a form of quasistatic evolution that cannot be rigorously justified
{except in the} case of a seasonal model that separates growth and stochastic dispersion. 
Fortunately, our numerical results in a number cases other than the seasonal model suggest the broader
applicability of this result. We further suggest a scheme (plotting the second moment as a function of
the mean) to test the validity of our proposed mechanism.

Several aspects of the model that require further study:
While the quasistatic arguments lead to a shape exponent of $\gamma =1+2D/\sigma^2$,
this is not the observed exponent even in the mean-field limit with a finite $\mu$. 
The complexities of an evolving probability distribution invalidate the quasistatic result.
 It may be possible that a more delicate treatment of the Fokker-Planck equation can
 help identify an effective Richards exponent.
The study of growth laws, or even statics, for the seascape Fisher equation in finite
dimensions is also interesting, and may have connections to directed percolation~\cite{korolev2010genetic}
and directed polymers in random media \cite{ginelli2003multiplicative,mangioni2000nonequilibrium}.

More complex connectivities are appropriate to social networks and spreading of contagions.  
Considering how the Richards equation has seen a recent rebirth in epidemiology and pandemic forecasting,
 it would be interesting to see whether implementing seascape growth on such human networks would 
  generate relevant results.  In particular, we may explore the role of long-range interactions 
  may be important to Richards growth~\cite{mombach2002mean,martinez2008continuous}.  
  
  While an experimental model can have complicating features, a  bacterial population in a well mixed fluid may 
  present a controllable realization of the mean-field model. Perhaps local growth rates can be varied
  stochastically by application of randomized light sources. The prediction of the model is that upon 
  increasing the strength of reproductive noise, the overall growth curve will change from logistic
  to Richards to Gompertz.

 
\section*{Acknowledgements}

This research was supported by James S. McDonnell Foundation Award No. 220020540 (B.O.-L.), as well as by NSF through Grants No.~DMR-1708280 and No.~PHY-2026995 (M.K.).

\section*{Contributions}
D.S. and B.O.-L. contributed equally to this manuscript.

\bibliography{SeaScape_Bib}{}

\clearpage

\appendix

\section{Numerical Recipies} \label{numerical_appendix}
We describe here the numerical tools  used to obtain the simulation results. 
The integration of stochastic equations is in general more difficult than the deterministic case. Deterministic ODEs can be treated with a variety of robust, accurate, and relatively simple methods, such as the fourth-order Runge-Kutta approach. In the stochastic case, 
we employ a method heavily inspired by that in Ref.~\cite{dornic2005integration}: 
The method we use involves splitting up the dynamics of the stochastic PDE 
\begin{equation}
    dy = (\mu y - a y^2 + D \nabla^2 y)dt + \sigma y dW\,,
\end{equation}
into pieces which can be solved individually. We first treat the stochastic piece,
\begin{equation}
    dy = \sigma y dW\,,
\end{equation}
whose solution, starting from any initial configuration $y(x,t)$, is 
a standard geometric Brownian motion
\begin{equation}
    y(x,t+dt) = y(x,t) \exp\left[-\frac{\sigma^2}{2}dt + \sigma dW(x,t)\right]\,,
\end{equation}
with $dW(x,t) \sim N(0,dt)$. 
Using the output of the stochastic step we then use a standard Euler 
update rule
\begin{equation}
    y(x,t+dt) = y(x,t) + dt\times D\nabla^2 y(x,t)\,,
\end{equation}
where $\nabla^2 y(x,t)$ is the appropriate discretization of the laplacian on the given lattice. For mean-field, $\nabla^2 y = \bar{y}-y$, while for a one dimensional lattice with unit spacing,  $\nabla^2 y(x,t) = y(x+1,t) + y(x-1,t)-2y(x,t)$. Finally, we take the output of this diffusion step and at each node apply a logistic growth, according to
\begin{equation}
    y(x,t) = \frac{\mu e^{\mu t} y(x,0)}{ay(x,0)(e^{\mu t} - 1) + \mu}\,.
\end{equation}
For a more complicated growth dynamics such as what is discussed in Section \ref{sec:universal}, we will resort to using a routine Runge-Kutta solver, as implemented in MATLAB's ode45 function, to circumvent the lack of an available analytic solution. MATLAB implementations of all routines discussed can be found on GitHub \footnote[2]{\url{https://github.com/dancewartz/RichardsGrowth}}.

\section{Seasonal growth dynamics} \label{appendix:twophase_moment}

To analyze the dynamics of the seasonal model in section~\ref{sec:seasonal}, it suffices to understand the dynamics of how the mean changes in a single growth phase.  At the start of the phase, we assume  a distribution of the form $\rho(y) \propto \hat\rho(y) = y^{-2-c_D}$ for $\Upsilon\leq y\leq \Lambda$, and 0 otherwise.  

The evolution of each node during the growth phase follows a deterministic logistic growth.  In particular, the population of a node after time $T_g$ is given by 
\begin{equation}
f(y; T_g) = \frac{y}{e^{-\mu T_g} + (1-e^{-\mu T_g})(y/y_f) },  
\end{equation}
where $y_f = \mu/a$ is the saturation for logistic growth.  
The average population at the end of the growth phase is then 
obtained as
\begin{equation}
\Exp{ y(T_g)} = \Exp{f(y; T_g)} = \frac{\int f(y; T_g) \hat\rho(y) dy}{\int \hat\rho(y) dy}.
\end{equation} 
Therefore, 
\begin{equation}
\Exp{ y(T_g)} = \frac{(1+c_D) e^{\mu T_g} M^{-c_D} }{{\Upsilon}^{-1-c_D} - \Lambda^{-1-c_D}}   \int_{M/\Lambda}^{M/\Upsilon} \frac{x^{c_D}}{x+1} dx,
\end{equation} 
where we define $M = y_f/(e^{\mu T_g} - 1)$.  It should be noted that there isn't a particularly convenient closed form for the integral as presented.  Moreover, since we assume $0<c_D<1$, the integral diverges as $(T_g)^{-c_D}$ for small times, meaning that it is sometimes more convenient to split the integral as $\int x^{c_D}/(x+1) = \int x^{c_D -1} - \int x^{c_D - 1}/(x+1)$.  For example, in the special case of $\Lambda \to \infty$, then this manipulation would become necessary, since the second integral would take on a term of $O(1)$ in time.  This leads to an anomalous growth term, where $\Exp{y(t)} - \Exp{y(0)} \propto t^{c_D}$, which emphasizes  the importance of the upper cutoff.  

If $\Lambda$ is finite, and we only care about  first order in time, then we indeed can just take the expectation of both sides of the logistic growth equation and find 
\begin{equation}
\frac{\Exp{y(T_g)} - \Exp{y(0)}}{T_g} = \mu \Exp{y(0)} - a \Exp{y(0)^2} + O(T_g).
\end{equation} 
Expanding the relevant integrals gives
\begin{alignat}{1}
\frac{\Delta \Exp{y}}{\Delta t} =\hspace{1mm}& \mu \Upsilon \frac{c_D+1}{c_D} \frac{1 - (\Upsilon/\Lambda)^{c_D}}{1 - (\Upsilon/\Lambda)^{1+c_D}} \notag \\
& + a \frac{c_D+1}{c_D-1} \frac{\Lambda^{1-c_D}}{{\Upsilon}^{-1-c_D}} \frac{1 - (\Upsilon/\Lambda)^{1-c_D}}{1 - (\Upsilon/\Lambda)^{1+c_D}}.  
\end{alignat} 
Since $1-c_D >0$ and $\Lambda \gg \Upsilon$, we can trivially expand this out as 
\begin{alignat}{1}
\frac{\Delta \Exp{y}}{\Delta t} =\hspace{1mm}& \mu \Upsilon \frac{c_D+1}{c_D} \left(1 - (\Upsilon/\Lambda)^{c_D}\right) \notag \\
& - a \frac{1+c_D}{1-c_D} \Lambda^{1-c_D}{\Upsilon}^{1+c_D} \left(1 - (\Upsilon/\Lambda)^{1-c_D}\right) \notag \\
& + O\left( (\Upsilon/\Lambda)^{1+c_D } \right),
\end{alignat}
which, if we neglect the higher order terms, returns us to Eq.~\eqref{eqn:DeterministicRichards}.  Therefore, a Richards-like growth law is retrieved.  

If we are interested in the limit of small $c_D$, notice how the first term of the expansion has a term like $(1-x^{c_D})/c_D$.  Therefore, the limit becomes 
\begin{alignat}{1}
\lim_{c_D \to 0^+} \frac{\Delta \Exp{y}}{\Delta t} =\hspace{1mm}& \lim_{c_D \to 0^+} \left(\mu \Exp{y} - a\Exp{y^2} \right) \notag \\
= &-\mu \Upsilon\frac{\log(\Upsilon/\Lambda)}{1-\Upsilon/\Lambda} - a \Lambda \Upsilon,
\end{alignat}
which is difficult to rewrite in terms of $\Exp{y}$ alone.


\section{Moments of Seascape Dispersion} \label{appendix:seascapediff_moment}
while not directly relevant to the study of scaling laws, for completeness we include a study of the dynamics of the moments, which can be used to determine relevant timescales for simulation.  
Let's consider a system under seascape noise and dispersion, much like the exploratory phase of the seasonal model of Sec.~\ref{sec:seasonal}, with 
\begin{equation}
dy = D(\bar{y} - y) dt + \sigma y dW\,.
\end{equation}
By applying It\^{o}'s lemma and taking the expectation of both sides, we get a set of ordinary differential equations of the form 
\begin{equation}
\partial_t \Exp{y^m} = A_m \Exp{y^{m-1}} + B_m \Exp{y^m}\,,
\end{equation}
with $A_m = mD\bar{y}$ and $B_m = m[ (m-1) \sigma^2/2 -D]$.  Note that each moment $m$ depends only on itself and the moment $m-1$, meaning that this is a solvable system.  For example, because $\Exp{y(t)} = \bar{y}$ for all time, this implies that the second moment is given by 
\begin{equation}
\Exp{y^2(t)} = \left(h_2 + \frac{2D{\bar{y}}^2}{\sigma^2 - 2D} \right) e^{(\sigma^2 - 2D) t}  - \frac{2D{\bar{y}}^2}{\sigma^2 - 2D}\,, 
\end{equation}
where $h_m \equiv \Exp{y^m(0)}$, and assuming generic values of $\sigma$ and $D$ (such that the denominator is never 0).  Notably, for $c_D = 2D/\sigma^2 < 1$, this means that the second moment grows exponentially.  Given that the $(m-1)$'th moment is known, the $m$'th moment is given by 
\begin{equation}
 \Exp{y^m} e^{-B_m t} = h_m  + A_m \int_{0}^t e^{-B_m s} \Exp{y^{m-1}(s)} ds.
\end{equation}
Assuming generic values of $D$ and $\sigma^2$, this recursion relation is solved by something of the form 
\begin{equation}
 \Exp{y^m} = \sum_{k=0}^m C_{m, k}  e^{B_m t}.
\end{equation}
The coefficients $C_{m,k}$ can be solved for exactly using inductive methods, though such precision is mostly of specialized use.  More importantly, this relation means that each moment above $m=1$ experiences leading-order exponential growth, with a growth rate of $m[ (m-1) \sigma^2/2 -D] > 0$ since $c_D < 1$.  Using this information, we can reasonably choose a befitting timescale for the exploration phase.


\end{document}